\documentclass[epj]{svjour}
\usepackage{graphics}
\newcommand{\naitl}{\mbox{NaI(Tl)} }
\newcommand{\up}[1]{$^{#1}$}
\newcommand{\down}[1]{$_{#1}$}
\newcommand{\Kthree}{\up{39}{K} }
\newcommand{\Kfour}{\up{40}{K} }
\newcommand{\Knat}{\textsuperscript{nat}{K} }

\newcommand{\keVee}{\mbox{keV\textsubscript{ee}} }
\newcommand{\kevee}{keV\textsubscript{ee} }

\newcommand{\pop}{Proof-of-Principle }
\newcommand{\spop}{SABRE-PoP }

\begin{document}
\title{The SABRE project and the SABRE Proof-of-Principle}
\author{M.~Antonello\inst{1}, E.~Barberio\inst{2}, T.~Baroncelli\inst{2}, J.~Benziger\inst{3}, L.~J.~Bignell\inst{4}, I.~Bolognino\inst{1,5}, F.~Calaprice\inst{6}, S.~Copello\inst{7,8}, D.~D'Angelo\inst{1,5}, G.~D'Imperio\inst{9}, I.~Dafinei\inst{9}, G.~Di Carlo\inst{7}, M.~Diemoz\inst{9}, A.~Di Ludovico\inst{6}, W.~Dix\inst{2}, A.~R.~Duffy\inst{10,11}, F.~Froborg\inst{12},  G.~K.~Giovanetti\inst{6}, E.~Hoppe\inst{13}, A.~Ianni\inst{7}, L.~Ioannucci\inst{7}, S.~Krishnan\inst{11}, G.~J.~Lane\inst{4}, I.~Mahmood\inst{2}, A.~Mariani\inst{8}, M.~Mastrodicasa\inst{9,14}, P.~Montini\inst{9,14}\thanks{\emph{Present address:} Dipartimento di Matematica e Fisica Universit{\`a} di Roma Tre I-00146 Roma and INFN Sezione di Roma Tre - 00146, Italy}, J.~Mould\inst{10,11}, F.~Nuti\inst{2}, D.~Orlandi\inst{7}, M.~Paris\inst{7}, V.~Pettinacci\inst{9}, L.~Pietrofaccia\inst{6}, D.~Prokopovic\inst{16}, S.~Rahatlou\inst{9,14}, N.~Rossi\inst{9}, A.~Sarbutt\inst{16}, E.~Shields\inst{6}, M.~J.~Souza\inst{6}, A.~E.~Stuchbery\inst{4}, B.~Suerfu\inst{6}, C.~Tomei\inst{9}, V.~Toso\inst{1,5}, P.~Urquijo\inst{2}, C.~Vignoli\inst{7}, M.~Wada\inst{6}, A.~Wallner\inst{4}, A.~G.~Williams\inst{15}, J.~Xu\inst{6}
}                     
\institute{INFN - Sezione di Milano, Milano I-20133, Italy \and 
School of Physics, The University of Melbourne, Melbourne, VIC 3010, Australia \and 
Chemical Engineering Department, Princeton University, Princeton, NJ 08544, USA \and 
Department of Nuclear Physics, The Australian National University, Canberra, ACT 2601, Australia \and 
Dipartimento di Fisica, Universit{\`a} degli Studi di Milano, Milano I-20133, Italy \and 
Physics Department, Princeton University, Princeton, NJ 08544, USA \and 
INFN - Laboratori Nazionali del Gran Sasso, Assergi (L'Aquila) I-67100, Italy \and 
INFN - Gran Sasso Science Institute, L'Aquila I-67100, Italy \and 
INFN - Sezione di Roma, Roma I-00185, Italy \and 
ARC Centre of Excellence for All-Sky Astrophysics (CAASTRO), Australia \and 
Centre for Astrophysics and Supercomputing, Swinburne University of Technology, PO Box 218, Hawthorn, Victoria 3122, Australia \and 
Imperial College London, High Energy Physics, Blackett Laboratory, London SW7 2BZ, United Kingdom \and 
Pacific Northwest National Laboratory, 902 Battelle Boulevard, Richland, WA 99352, USA \and 
Dipartimento di Fisica, Sapienza Universit{\`a} di Roma, Roma I-00185, Italy \and 
The University of Adelaide, Adelaide, South Australia, 5005 Australia  \and 
Australian Nuclear Science and Technology Organization, Lucas Heights, NSW 2234, Australia 
}
\date{Received: date / Revised version: date}
\abstract{
SABRE aims to directly measure the annual modulation of the dark matter interaction rate with NaI(Tl) crystals. A modulation compatible with the standard hypothesis in which our Galaxy is embedded in a dark matter halo has been measured by the DAMA experiment in the same target material. Other direct detection experiments, using different target materials, seem to exclude the interpretation of such modulation in the simplest scenario of WIMP-nucleon elastic scattering. The SABRE experiment aims to carry out an independent search with sufficient sensitivity to confirm or refute the DAMA claim. The SABRE concept and goal is to obtain a background rate of the order of 0.1 cpd/kg/\keVee in the energy region of interest. This challenging goal is achievable by operating high-purity crystals inside a liquid scintillator veto for active background rejection. In addition, twin detectors will be located in the northern and southern hemispheres to identify possible contributions to the modulation from seasonal or site-related effects. The SABRE project includes an initial Proof-of-Principle phase at LNGS (Italy), to assess the radio-purity of the crystals and the efficiency of the liquid scintillator veto. This paper describes the general concept of SABRE and the expected sensitivity to WIMP annual modulation. 
} 
\titlerunning{The SABRE project and the SABRE Proof-of-Principle}
\authorrunning{M.~Antonello et al.}
\maketitle
%
\section{Motivation}  
\label{sec:1}
For decades, direct detection experiments have been searching for interactions of dark matter candidates, generally Weakly Interacting Massive Particles (WIMPs)~\cite{PhysRevLett.39.165}, with a target material.
A net flux of dark matter through terrestrial detectors is expected assuming that the solar system moves through a dispersion dominated dark matter halo surrounding our galaxy. Due to the Earth's orbital motion around the Sun, the predicted interaction rate undergoes an annual modulation with a characteristic phase~\cite{PhysRevD.37.3388}. 
The DAMA experiment~(short for DAMA/NaI and DAMA/LIBRA) has observed a clear annual modulation exploiting \naitl crystals at Laboratori Nazionali del Gran Sasso (LNGS) in Italy. This is a model-indepen\-dent finding which satisfies the criteria for a WIMP induced signal~\cite{dama2013}. Recently, the DAMA collaboration has released the first results from their Phase-2 experiment~\cite{DAMAPhase2}, confirming the evidence of a signal that meets all the requirements of a model-independent dark matter annual modulation signature at $12.9\,\sigma$ significance. 
Results from several other experiments, when interpreted in the standard WIMP galactic halo hypothesis, seem to exclude the interpretation of the DAMA signal as due to spin-independent WIMPs nuclear scattering. Currently, the best sensitivity in the relevant mass region is reported by the XENON1T experiment~\cite{xenon1t}. However, given the different target materials of these experiments, a model-independent comparison of the results is not possible.

The Sodium-iodide with Active Background REjection (SABRE) experiment aims to perform a measurement of the dark matter annual modulation with \naitl crystals, but with a background lower than that of the DAMA experiment. 
The experiment will consist of twin detectors located in both the Northern and Southern hemispheres to disentangle possible seasonal or site effects from the dark matter modulation.  
The dual site is a unique feature of SABRE, with respect to other NaI(Tl) dark matter searches currently running: the COSINE-100 experiment~\cite{Adhikari2018} at the YangYang Laboratory in South Korea and the ANAIS experiment~\cite{Amare:2017kht} at the Canfranc Laboratory in Spain. These experiments, even after several years of operation, might not be able to resolve all possible scenarios in interpreting the DAMA signal as a dark matter signature, given their background level which is about 2-3 times higher than the DAMA one. 
On the other hand, if the modulation is observed, a precision measurement by SABRE might offer insights on how dark matter particles interact with ordinary matter and perhaps on their density and velocity in the galactic halo~\cite{Gluscevic2015,FreeseLisantiSavage}. 
The hypothesis of a focusing effect of the dark matter wind due to the gravitational potential of the Sun could also be investigated~\cite{Lee2013}. 
\section{The SABRE concept}
\label{sec:2}
The SABRE experiment uses \naitl scintillating crystals for dark matter detection and focuses on the achievement of a background of the order of 0.1 cpd/kg/\kevee in the energy region of interest for a dark matter search (order of \kevee for \naitl detectors). 
 
The substantial background reduction anticipated by SABRE, namely one order of magnitude below the level reached by the DAMA experiment and not yet matched by any other \naitl experiment to date, will be achieved via ultra-high crystal radiopurity coupled with active rejection through a liquid scintillator veto. 
Both features will be tested in an initial \pop phase, denoted \spop. 

The \spop setup is presently under construction at LNGS under a rock coverage equivalent to 3600 m of water. \spop is expected to run prior to the deployment of the full-scale experiment. This will consist of two twin detectors: SABRE-North at LNGS and SABRE-South at the Stawell Underground Physics Laboratory~(SUPL) in the Southern hemisphere (see Sect.~\ref{sec:twins}).

The following sections introduce the main ideas of the SABRE project and describe the Research \& Development carried out so far by the SABRE Collaboration. 

\subsection{NaI(Tl) detectors}
\label{sec:crystal}
A crucial part of the SABRE effort is the development of high-purity \naitl crystals. The background from decays due to radioactive impurities in the crystals is indeed the hardest to suppress. These processes can produce energy signals of a few \kevee, compatible with the expected recoil energies due to dark matter interactions and within the energy range where the DAMA modulation is observed (2-6~\kevee~\cite{dama2013}).
Among the potential background sources, \Kfour\ is the most dangerous since one of its decay channels causes a 3~keV Auger/X-ray emission. Other contaminants to limit include \up{238}U, \up{232}Th, \up{87}Rb. Worrisome contributions also come from \up{210}Pb, \up{3}H and cosmogenic activated isotopes. For a study of the background contribution from internal radioactivity of the crystals we refer to the Monte Carlo simulations performed for the \pop phase~\cite{SABREPoPMC}.

To keep the intrinsic contaminants at a very low level, SABRE has developed a method to obtain ultra-pure NaI powder and a clean procedure to grow crystals.
Princeton University and industrial partner Sigma-Aldrich~\cite{SigmaAldrich} have managed to produce ultra-high purity NaI powder, so-called Astro Grade powder, with potassium levels consistently lower than $\sim$ 10 ppb.
The content of \up{238}U and \up{232}Th in Astro Grade powder has been measured by ICP-MS~\cite{PNNLMeas} obtaining for both the isotopes the upper limit of 1~ppt. 
The \naitl crystals are then grown by Radiation Monitoring Devices, Inc. (RMD)~\cite{RMD}. They use the vertical Bridgman-Stockbarger technique~\cite{Bridgman1925}, where the powder is placed inside a sealed ampoule. The sealed environment reduces the possibility of contamination of the material during the growth phase. \\ To determine the best ampoule composition, several materials were prepared with different cleaning procedures. 
Crystals were grown from Astro Grade powder following these procedures and tested for radioactive contaminations with Inductively Coupled Plasma Mass Spectrometry~(ICP-MS).  
The optimum ampoule composition, together with a precision cleaning, showed no increase of the impurity levels inside the crystal with respect to the starting powder. 
At the end of 2015, the optimal procedure was used to grow a 2-kg crystal with average \Kthree\ level of $9\pm1$~ppb~\cite{PNNLMeas,ARNQUIST201715,SeastarMeas} and with \up{87}Rb upper limit of 0.1~ppb measured by ICP-MS~\cite{SeastarMeas}. 
As a comparison, DAMA reports an average of 13~ppb \Knat\ in their crystals (equivalent to 12.1~ppb of \Kthree), \up{238}U and \up{232}Th content below 0.01~ppb, and an upper limit on the content of \up{87}Rb ($<$0.35~ppb)~\cite{damaapp}.

The crystal purity can only be partially evaluated by ICP-MS or Accelerator Mass Spectrometry (AMS), a mass spectrometry that involves accelerating the ions to extraordinarily high kinetic energies. Both techniques are only sensitive to primordial parents. Direct counting of the intrinsic radioactivity by operating the crystal as a scintillator
is still the most accurate method and can also detect cosmogenic activation that occurred prior to installation of the crystal underground.
The \spop phase is in preparation at LNGS to assess the SABRE crystal purity by directly operating the crystal as a scintillator inside an active veto in an underground setup.
RMD is currently growing cylindrical \naitl crystals for a mass of about $5$~kg for each crystal. The first of these crystals will be tested within the PoP setup. 

In the SABRE design, two PMTs are directly coupled onto each flat end of the crystal, while its curved side is wrapped with reflector for better light collection efficiency.
We anticipate using PMTs with high quantum efficiency, very low background, and low dark counts in order to access energies below 2 \kevee. 
The detector assembly is placed inside a high-purity,  air- and light-tight copper enclosure. 
The technical design of the detector module developed for the \spop phase is described in Sect.~\ref{PoPdesign}.

\subsection{Active background rejection system}
\label{sec:rejection}
SABRE aims to further reduce the background level by placing an array of detector modules (\naitl crystals coupled to PMTs) inside a liquid scintillator veto. This system consists of a vessel filled with scintillating fluid and equipped with PMTs on its walls. Background events from both intrinsic or external sources are likely to deposit energy also in the veto or in adjacent crystals. 
Events with simultaneous energy depositions in the veto and the crystal can be identified as background. 
For instance, the dangerous 3~\kevee signal following the electron capture of \Kfour\ in the \naitl crystal can be rejected by detecting the coincident  1.46~MeV gamma signal in the surrounding material. Experiments such as DAMA and ANAIS~\cite{Amare:2017kht} can only reject events with coincident energy deposition in neighbouring crystals, achieving a partial suppression of this background. However, the SABRE strategy (adopted also by the COSINE~\cite{Adhikari2018} experiment) with 4$\pi$ active volume around the crystal matrix, has the potential to identify and suppress the \Kfour\ background with higher efficiency.
We anticipate using an organic scintillator mixture based on pseudocumene and high quantum efficiency PMTs. This combination will be tested within the \spop setup, as described in Sect.~\ref{PoPdesign}. 
SABRE-South is exploring an alternative liquid scintillator based on linear alkyl benzene (LAB) solvent, due to its less stringent safety handling constraints (higher flash point).

\subsection{Twin detectors}
\label{sec:twin}
\label{sec:twins}
A unique feature of SABRE is to adopt twin detectors located in the North and South hemispheres. The annual modulation of the experimental rate induced by dark matter interaction is expected to have the same phase in both hemispheres, given its galactic origin. In contrast, an annual modulation of the rate due to seasonal or site effects would be characterized by a different phase and amplitude in the two detectors.  
The SABRE twin detectors will be placed at LNGS, Italy, and at the Stawell Underground Physics Laboratory (SUPL) in
Australia. SUPL will be the first underground laboratory in the southern hemisphere and is located $240$ km north-west of Melbourne. A section of an active gold mine is being converted into the laboratory.  The chosen site is 1025 m deep with a flat overburden, corresponding to a $\sim $3000 m water equivalent depth, which is similar to LNGS.

\section{\spop Technical Design}
\label{PoPdesign}
The \spop phase has the goal of assessing the crystal purity, the effectiveness of the active background rejection system, and the overall background level. 
This phase will run with a single high-purity cylindrical \naitl crystal, for a total mass of $\sim5$-kg, which is currently being grown by RMD from Astro Grade powder following the procedure described above. The crystal will be wrapped with Polytetrafluoroethylene (PTFE) as reflector and coupled at each end to 3-inch Hamamatsu R11065-20 PMTs.
These PMTs are specifically developed for low-background experiments as they have high quantum efficiency of 30-35\% at 420~nm and low intrinsic radioactivity~\cite{pmtback_2015}. 

\begin{figure}
\resizebox{0.5\textwidth}{!}{%
  \includegraphics{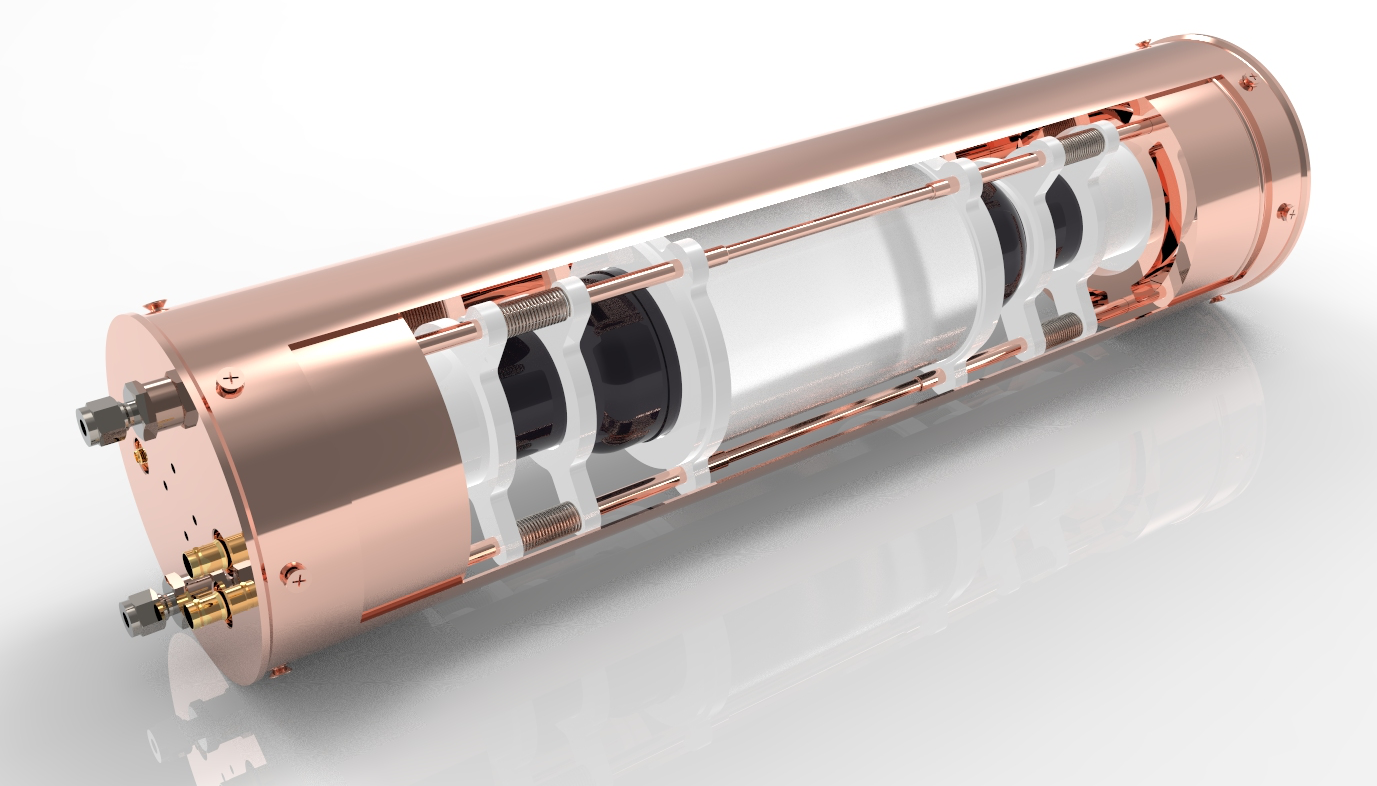}
}
\caption{Rendering of the detector module for the \spop phase. The crystal is placed in a copper enclosure; the crystal wrapping is not shown. Two PMTs (black body and blue window) are coupled on each end of the crystal. PFTE holders (white) and copper rods keep the crystal and the PMTs in position.}
   \label{fig:enclosure}
\end{figure}

The assembly will be sealed in a copper enclosure, as shown in Fig.~\ref{fig:enclosure}. 
The cylinder, the end-caps and the support rods are made out of C10100 copper by Alca Technology~\cite{Alca}. The holders for the crystal and the PMTs are made out of high-purity PTFE. Given that sodium-iodide is highly hygroscopic, all parts will be thermally treated prior to assembly to remove any residual water. After the sealing, the detector module will be flushed with dry and clean low Ar, low Kr (LAK) N\down{2} gas to avoid humidity and radon.  
The cylindrical part of the enclosure has a diameter of 14.6~cm, a height of 58~cm and a thickness of 2~mm. It is closed by a shaft seal with Viton\textregistered\ O-rings. 
The enclosure sits vertically and the top end-cap is furnished with bulkhead feedthroughs for the PMT high voltage and signal cables and for teflon tubes. The latter will allow flushing of the inner volume with dry LAK nitrogen gas. \\
\indent The copper enclosure will be placed inside a $1.3$~m (diameter) $\times 1.5$~m (length) cylindrical vessel filled with about two tons of pseudocumene and instrumented with ten 8-inch PMTs. The vessel is made from low-radioactivity stainless steel with access via a 60~cm diameter 
flange at the top. 
The vessel's inner surface is coated with Ethylene tetrafluoroethylene (ETFE) to prevent scintillator degradation due to direct contact with the stainless steel. 
A Lumirror\textsuperscript{TM} liner with 95\% reflectance above 400~nm is applied inside to improve light reflection.
Each flat end of the vessel hosts five Hamamatsu R5912-100 PMTs with 35\% quantum efficiency at 390 nm.
The liquid scintillator is distilled pseudocumene (from the Borexino facility~\cite{ALIMONTI200958}) doped with 3 g/l of PPO (2,5)-diphenyloxazole, which acts as wavelength shifter.
The expected light yield is 0.22~photoelectrons/\kevee~\cite{EmilyThesis}. \\
\indent The detector and the fluid handling system design foresee a slow control system that will ensure precise and stable control over all operational parameters such as temperature, pressure, and radon emanation, with the double goal of ensuring security and control of all variables relevant for the physics analysis.   \\
\indent To allow the insertion of the detector module, through the top flange of the vessel, a 2~mm thick copper tube (16~cm diameter and 121~cm height) is connected to the top cover plate of the vessel and provides a dry volume inside the steel vessel.
The detector module will be inserted in this tube without any contact of the scintillator with air or the outside environment during the operation.
The inner volume of the tube is flushed with dry and clean N\down{2}-gas through teflon tubes running into the small space between the copper tube and the enclosure. 
This flushing serves as a safety blanket against moisture, radon gas or other background sources. 
An additional teflon tube allows for the insertion of wire-mounted calibration sources. 
To ensure a precise insertion of the copper tube as well as the detector module, a removable frame holding a motorized pulley will be mounted atop the vessel. \\
\begin{figure}
\resizebox{0.5\textwidth}{!}{%
        \includegraphics{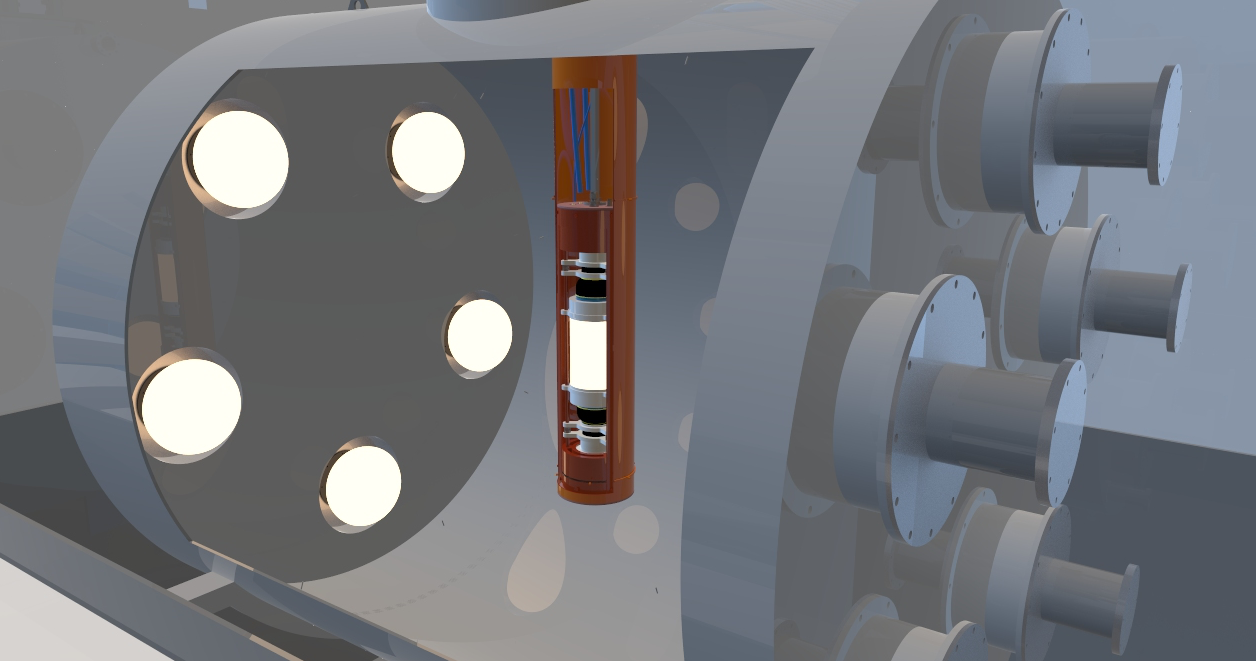}
        }
\resizebox{0.5\textwidth}{!}{%
        \includegraphics{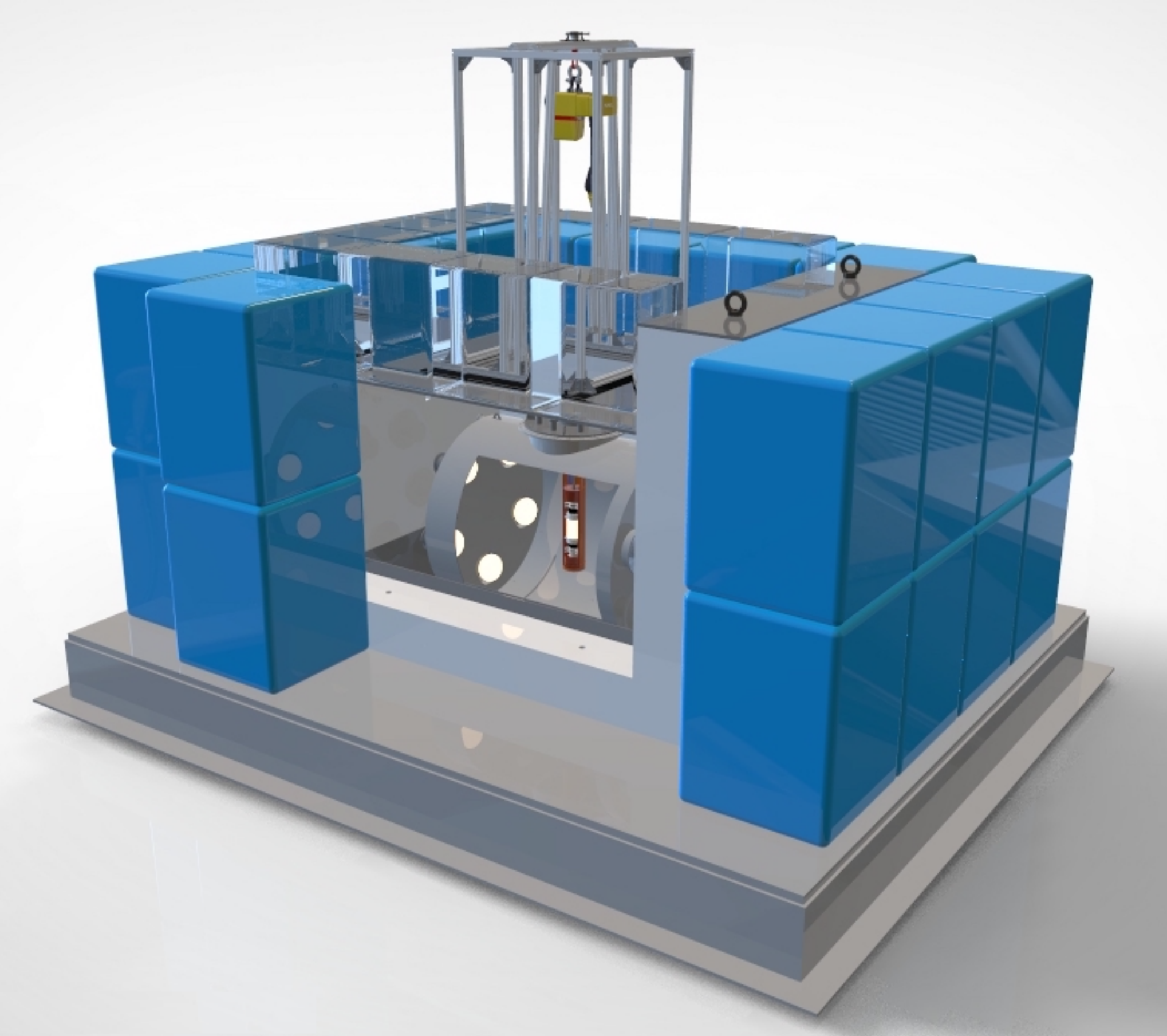}
        }
	\caption{Top: the \spop insertion system. A thin copper tube is connected to the top cover plate of the vessel to provide a dedicated dry volume for the detector module. Bottom: a removable frame is mounted at the top of the vessel to lower the detector module with a motorized pulley into the tube. The vessel is surrounded on each side by polyethylene. On the bottom, the detector is isolated by lead layers, while the sides and the roof are shielded by water tanks.}
	\label{fig:shielding}
\end{figure}
\indent To further shield the setup from external radiation, the vessel is surrounded by several layers of passive material. The innermost layer is made of polyethylene, with 10~cm thickness on the top and bottom and at least 40~cm on the four sides. The polyethylene supports a 2~cm steel plate at the top, where the insertion system can be mounted. Water tanks are placed on top of the plate and on the sides of the shielding for a water thickness of 80~cm and 91~cm, respectively. The shielding castle is completed with a 15~cm lead floor. The technical design of the insertion system and of the passive shielding surrounding the vessel is shown in Figure~\ref{fig:shielding}. The polyethylene shell will be sealed, allowing the inner volume to be flushed with dry and clean N\down{2}-gas, to avoid radon gas which is present in the laboratory air. \\
\begin{figure}
\resizebox{0.49\textwidth}{!}{%
\includegraphics{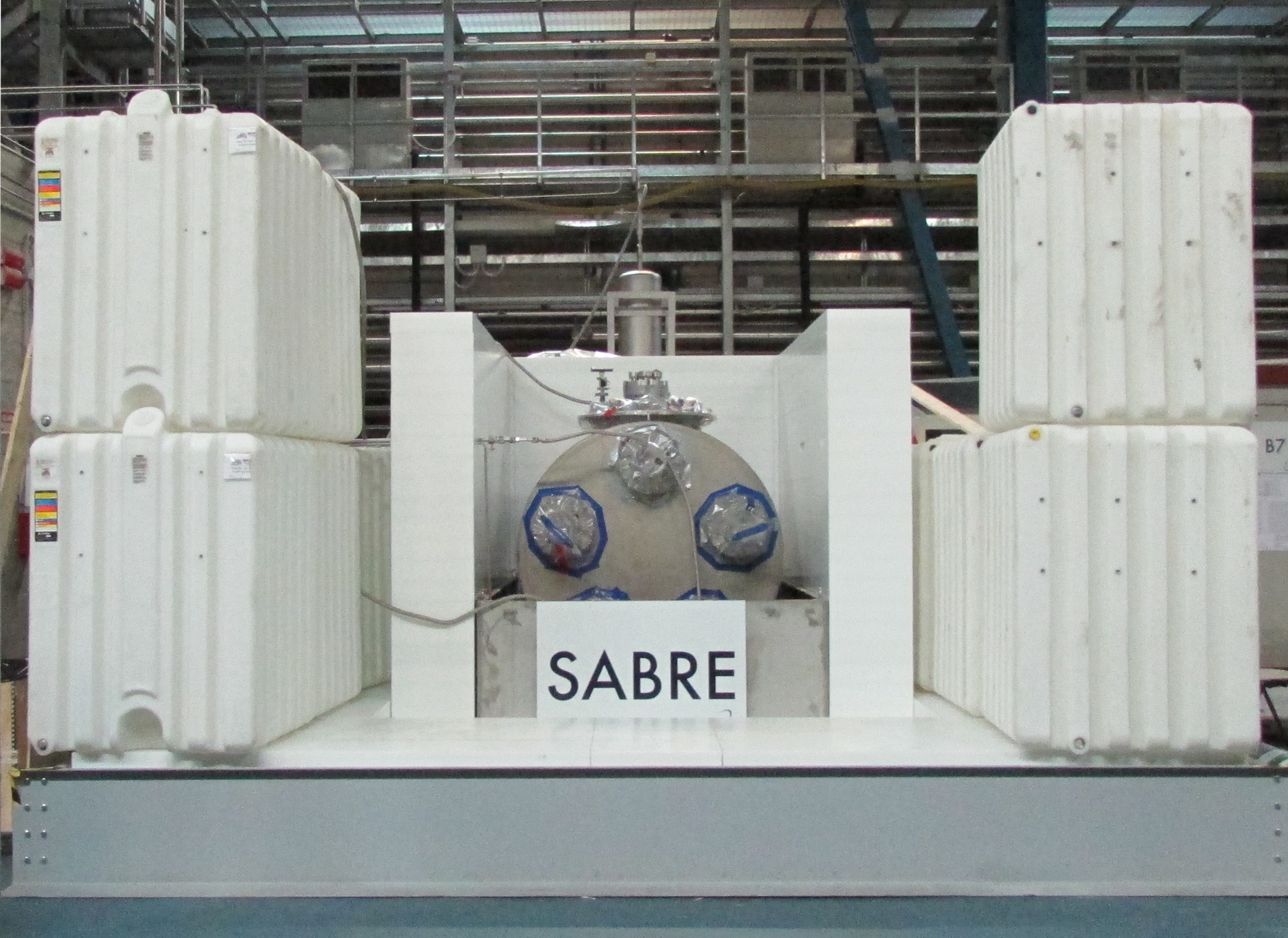}
}
	\caption{The \spop setup in LNGS Hall-C}
	\label{fig:PoPassembly}
\end{figure}
\indent The installation of the \pop setup is currently ongoing at LNGS (see Fig.~\ref{fig:PoPassembly}) and data taking is foreseen in 2019.
\section{Sensitivity to the annual modulation}
In this section we study the sensitivity reach of the full-scale SABRE experiment, that will build on the experience and results of the \pop phase. The sensitivity has been studied as a function of the experimental characteristics, such as target mass, measuring time and background level. The results reported in the following assume a total mass of 50~kg of NaI(Tl) crystals over three years of data collection. We use two different approaches: at first, we study the statistical significance to confirm or reject a signal's modulation with the amplitude reported by DAMA, regardless of any hypothesis on its origin. Later, we draw the standard 90\% C.L. sensitivity curve for SABRE to spin-independent WIMP-nucleon scattering and we show it in comparison with the allowed regions for the DAMA result under these assumptions. We refer to the modulation amplitude reported in~\cite{dama2013} for DAMA/NaI and DAMA/LIBRA Phase-I as ``the DAMA result" throughout the following.

Concerning the background level, we rely on Geant4-based~\cite{geant} Monte Carlo simulations performed for the \- \spop setup. For these simulations, we assumed the PoP crystal had the same radio-purity as the 2 kg test crystal described in section~\ref{sec:crystal}. Where no measurement was available for the crystal itself, the radio-purity of the Astro Grade powder or, if that was not measured, of the DAMA crystals~\cite{damaapp}, was assumed. We used the ACTIVIA~\cite{ACTIVIA} simulation software to calculate the activation levels of cosmogenic isotopes, assuming an exposure at sea level of about 1 year plus a transport by plane from USA to Italy. Concerning the other detector components included in the simulation (crystal and veto PMTs, copper enclosures, vessel, liquid scintillator, etc...), we used radioactive contamination levels either directly measured by the SABRE collaboration or 
by other experiments for the same material or item. 

We find that the radioactive contamination of the crystal gives the most significant contribution to the background in the energy region of interest for a dark matter search. 
About 20\% of the background is accounted for by \up{87}Rb, U and Th contaminations, whose values used in the simulation are actually upper limits in the NaI powder. Background from cosmogenically-activated isotopes is dominated by \up{3}H, that accounts alone for about 50\% of the total background. Cosmogenic activation is conservatively estimated after transport by place and only 180 days of underground storage. Transport by ship and longer storage time could actually lead to significant reduction of the activity.
The predicted background level in the energy interval [2-6]~\kevee for the \spop setup is 0.36~cpd/kg/\kevee. This is obtained after rejecting all the events accompanied by an energy deposit above a $100$~keV threshold in the liquid scintillator veto. Further details on the background estimates and related simulations can be found in~\cite{EmilyThesis,SABREPoPMC}. Even though the predicted background level for the \spop is higher than the goal we anticipate for the full scale experiment, we conservatively evaluate the SABRE sensitivity using 0.36~cpd/kg/\kevee instead of 0.1~cpd/kg/\kevee.
 
In order to estimate the sensitivity of SABRE to the DAMA signal, irrespectively of any assumption on its interpretation, we use toy MC simulations of the experimental rate. We generate 50k data sets that include the expected background and a modulating signal with an amplitude of 0.011 cpd/kg/\kevee, which is compatible with both Phase-1 and Phase-2 results published by DAMA in the [2-6]~\kevee energy range~\cite{dama2013,DAMAPhase2}. Fitting each data set with a sinusoidal function with a period fixed to one year, we obtain a gaussian distribution of the fit amplitudes around the injected value. 90\% of the experiments show a modulation amplitude with a significance of $5.5\,\sigma$. We also generate 50k data sets under a background-only hypothesis. Fitting with the same function, the distribution of the amplitudes centers at zero. 90\% of the simulated experiments show a modulation amplitude incompatible with the DAMA signal at $5\,\sigma$.   

Next, we evaluate the sensitivity of SABRE to spin-independent WIMP nuclear scattering.
The expected dark matter interaction rate on a NaI target has been evaluated as a function of time using the expression given in~\cite{FreeseLisantiSavage}. This assumes the standard WIMP halo model and a spin-independent WIMP nucleon interaction. We also assume an average Earth velocity of 232 km/s, a Galaxy escape velocity of 
544 km/s, and that the WIMP velocity follows a Maxwellian distribution with most probable speed 
of 220 km/s. Concerning the quenching factor of nuclear recoils on sodium, we applied the energy dependent measurements from~\cite{Xu2015_QF}. 
As for iodine we assumed the value of 0.09, also used by DAMA~\cite{dama1996} with an error of 0.01 assuming the analysis described in~\cite{tretyak}. 
The detection efficiency and resolution were set at the values reported by the DAMA collaboration~\cite{damaapp} for their detectors.

\begin{figure}
\resizebox{0.5\textwidth}{!}{%
	\includegraphics{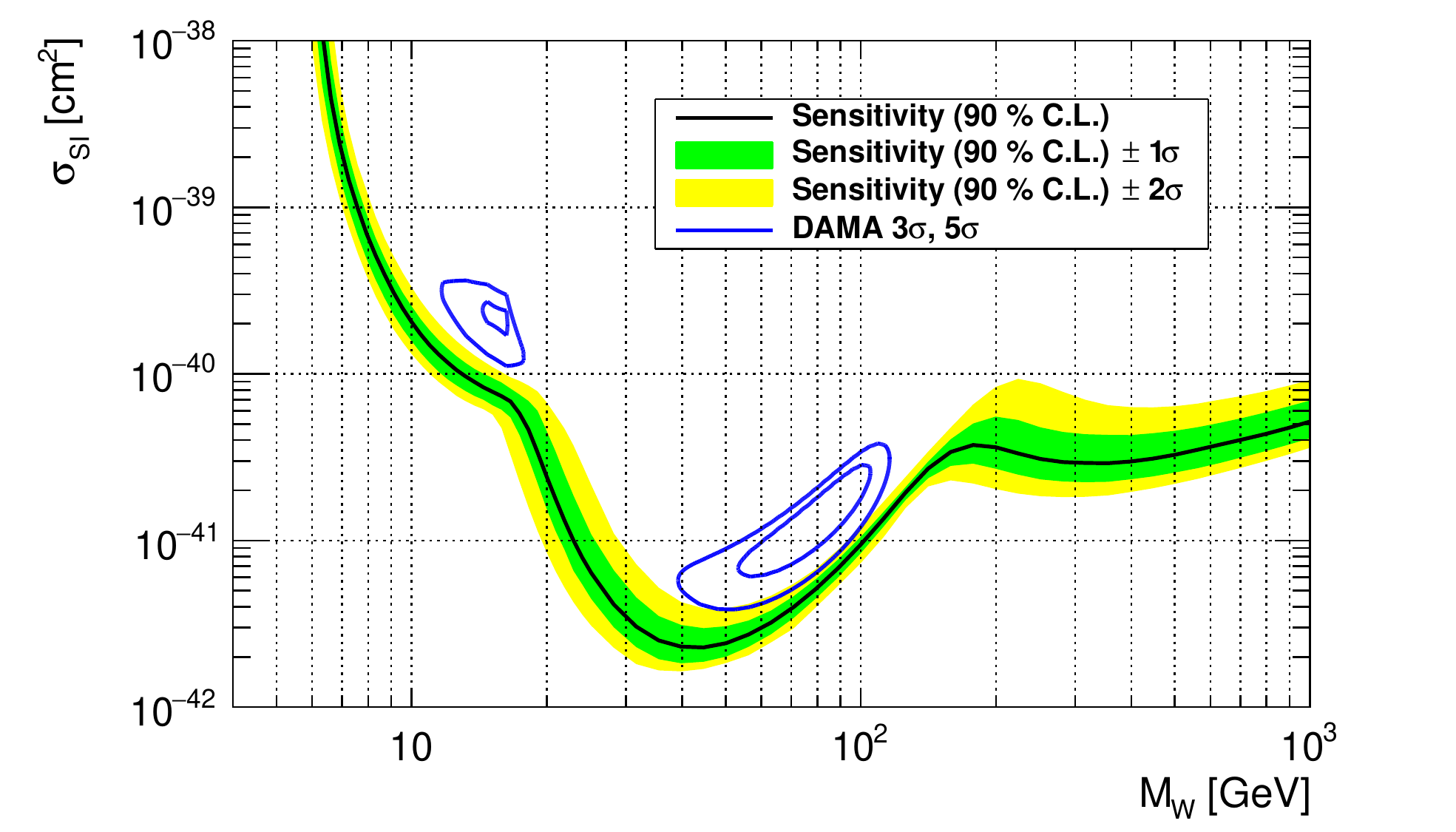}
	}
	\caption{SABRE sensitivity for a total mass of 50~kg of NaI(Tl) crystals over three years of data collection, with uncertainty bands that cover different modeling of efficiency and energy resolution. The blue curves represent the 3 and 5 sigma confidence regions we obtained interpreting the DAMA Phase-1 results~\cite{dama2013,dama2013Rev} in the framework of the standard WIMP model, as described in~\cite{Savage09}.} 
	\label{fig:sensitivity}
\end{figure}

The sensitivity curve for SABRE was calculated as follows.
For several combinations of WIMP mass ($M_{W}$) between 1 GeV and 1000 GeV and cross section ($\sigma_{SI}$) between $10^{-42}$ cm$^2$ and $10^{-36}$ cm$^2$, we generate 1000 data sets binned in 30-day intervals, obtained with a Poisson extraction that accounts for the expected unmodulated rate plus the background from Monte Carlo. 
This process is performed by varying the energy from 2 to 6 keV in eight 0.5 keV bins. 
We fit each data set with a cosine function having a 1-yr period and we obtain a zero-centered gaussian distribution of amplitudes. We then sum over the eight bins the $\chi^2$ of the modulation amplitude function with respect to the non-modulating data set, considering the RMS of the respective gaussian distribution of amplitudes as the sigma of each bin. 
The resulting $\chi^2(M_{W}, \sigma_{SI})$ is cut at 90\% C.L. and the sensitivity curve is shown in Fig.~\ref{fig:sensitivity}.
We have added to the plot the 3- and 5-sigma confidence regions we obtained interpreting the DAMA Phase-1 results~\cite{dama2013,dama2013Rev} in the framework of the standard WIMP model, as described in~\cite{Savage09} under the same assumptions on WIMP interaction, halo model, quenching factor, energy resolution and detection efficiency. There are two caveat on this plot; the first is that, as stated, the comparison is model-dependent. The second is that the interpretation of DAMA Phase-2 results with a 1~\keVee energy threshold~\cite{DAMAPhase2} seems disfavoured in the framework of the standard WIMP model. Other theoretical explanations are being proposed~\cite{Baum:2018ekm,Kang:2018qvz,Herrero-Garcia:2018lga}. 

We have also produced similar plots using different assumptions for the sodium quenching factor, given that several independent measurements are available~\cite{dama1996,Xu2015_QF,QFtexas} and verified that the relative position of our sensitivity curve and the lower energy DAMA-allowed region is independent of this parameter.

We studied the impact on the sensitivity curve of varying the energy resolution, the detection efficiency, the sodium and the iodine quenching factors, and the background level within their uncertainties.
The systematics associated with the choice of the energy binning was found to be negligible around our choice of 0.5 keVee. All of the above parameters have been simultaneously varied by randomly sampling 1000 times from their expected distributions. We used each set of sampled values to calculate 1000 expected annually modulated daily rates, fit with a cosine function, build the distribution of sensitivity values as described above, and extract the 90\% C.L. sensitivity as the median of that distribution (black solid line) and the 1$\sigma$ (green) and 2$\sigma$ (yellow) regions of Fig.~\ref{fig:sensitivity}. 
The experiment is sensitive to spin-independent WIMP-nucleon scattering cross sections as small as $2\cdot 10^{-42}$~cm$^{2}$ for a WIMP mass of 40-50~GeV.

\section{Conclusions}
The SABRE experiment will investigate the expected annual modulation due to interaction of dark matter particles in the galactic halo on 
\naitl scintillating crystals. \\ SABRE aims to achieve a background of \mbox{0.1~cpd/kg/\kevee} via crystal purity and active rejection through a liquid scintillator veto. 
The experiment also adopts PMTs with high quantum-efficiency and low background, which are directly coupled to the crystals, to maximize the light collection efficiency and gain sensitivity for energies below 2 \kevee.  
SABRE foresees the installation of twin detectors underground at LNGS in Italy and at SUPL (Stawell Underground Physics Laboratory) in Australia. The dual location in opposite hemispheres allows the identification of any local and/or seasonal effects that could possibly contribute a modulation of the experimental rate. 
SABRE will either verify or refute, in a model-independent way, the long debated result of the DAMA experiment, as a result of SABRE being the highest sensitivity NaI(Tl) based experiment.
Under the assumption of a spin-independent WIMP-nucleus interaction, after three years of exposure and with a total mass of 50~kg, the experiment is expected to be sensitive to WIMP-nucleon scattering cross sections down to $2\cdot 10^{-42}$~cm$^{2}$ for a WIMP mass of 40-50 GeV.

\section*{Acknowledgements}
The SABRE program is supported by funding from INFN (Italy), NSF (USA), 
and ARC (Australia\footnote{Grants: LE170100162, LE16010080, DP170101675, LP150100075}). F.~Froborg has received funding from the European Union's Horizon 2020 
research and innovation programme under the Marie Sklodowska-Curie grant agreement No 703650.
We acknowledge the generous hospitality and constant support of the Laboratori Nazionali del Gran Sasso (Italy).
%
\bibliographystyle{spphys}
\bibliography{bibliography}

\end{document}